# A Method For Business Process Reengineering Based On Enterprise Ontology

Pedram Bahramnejad[1], Seyyed Mehran Sharafi[2], Akbar Nabiollahi[3]

[1]Department of Computer Engineering, Najafabad Branch, Islamic Azad University, Najafabad, Iran
[2]Department of Computer Engineering, Najafabad Branch, Islamic Azad University, Najafabad, Iran
[3]Department of Computer Engineering, Najafabad Branch, Islamic Azad University, Najafabad, Iran

## ABSTRACT

*Business Process Reengineering increases enterprise's chance to survive in competition among organizations , but failure rate among reengineering efforts is high, so new methods that decrease failure, are needed, in this paper a business process reengineering method is presented that uses Enterprise Ontology for modelling the current system and its goal is to improve analysing current system and decreasing failure rate of BPR, and cost and time of performing processes, In this method instead of just modelling processes, processes with their : interactions and relations, environment, staffs and customers will be modelled in enterprise ontology. Also in choosing processes for reengineering step, after choosing them, processes which, according to the enterprise ontology, has the most connection with the chosen ones, will also be chosen to reengineer, finally this method is implemented on a company and As-Is and To-Be processes are simulated and compared by ARIS tools, Report and Simulation Experiment*

## 1. INTRODUCTION

In 1970s the American auto industry faced with an attack from the Japanese automakers which shook their foundations. The Japanese were able to make high-quality cars at prices much cheaper than the American giants like Ford and Chrysler. Suddenly Ford and Chrysler realized there was something Japanese giants like Toyota and Nissan were doing different which was making them so very competitive. This danger of being displaced as the market leader from their very own home turf led to a severe introspection which resulted in many management paradigms like Six Sigma, TQM, and PIP[1] , Which were based on incremental changes in the organization, and could improve the business processes. In 1990s they knew that those methods could improve the business to some certain levels and something new and more fundamental is needed for taking business to next level, In such times Prof. Michael Hammer wrote his important article[2]. Hammer claimed that "…the major challenge for managers is to obliterate non-value adding work, rather than using technology for automating it" Similar views were expressed by Thomas

H. Davenport and J. Short in[3] Main idea was defined formally by Hammer and Champy in 1993 as "…the fundamental reconsideration and radical redesign of the organizational process, in order to achieve drastic improvement of current performance in cost, service and speed"[4] Guimaraes and Bond in[5] wrote " BPR aims at making these processes more competitive by improving quality, reducing costs, and shortening the product development cycle".

Reengineering assumes the current process is irrelevant, it doesn't work, it's broke, forget it, start over. Such a clean slate perspective enables the designers of business processes to disassociate themselves from today's process, and focus on a new process. We can say it is kind of like going to the future and asking yourself: What should the processes look like? What do customers want it to be like? What do employees want it to be like? How do high class organizations do it? What can we do with new technology[1]?

There are many different methods and approaches for performing BPR that will be introduced later but It is difficult to find a single approach exactly matched to a particular organization's needs, and the challenge is to know what method to use when and how to pull it off successfully such that bottom-line business results are achieved [1]. This paper tries to propose a Bpr method based on Enterprise Ontology and implementing it on a distributing segments of car's body company named "Barez Pakhsh" In kerman, IRAN, and comparing it with data of another Bpr method.

## 2. RELATED WORKS

Organizations use different Bpr methods, steps of two of these methods are presented here in order to have an understanding of a Bpr methodology.

### 2.1. Hammer and Champy Methodology

Hammer and champy's methodology for Bpr consists of six steps.
1. Introduction into business reengineering
2. Identification of business processes
3. Selection of business processes:
4. Understanding the selected business processes
5. Redesign of the selected business processes
6. Implementation of redesigned business processes: [4, 6].

### 2.2. Davenport methodology

1. Visioning and goal setting
2. Identification of business processes
3. Understand and measure
4. Information technology
5. Process prototype.
6. Implementation[3, 6].

### 2.3. Enterprise Ontology

In most Bpr steps, IT plays the main role and today without IT, Bpr cannot work properly, IT is one of the most important parts of Bpr but this doesn't mean that using IT alone can be as useful as Bpr. In most published papers concerning Bpr such as[7, 8], IT is the base of their methodology for redesigning and even evaluating the processes. In 2012 a paper published [9] that formed the foundation of this paper, it proposed a method for making knowledge maps and knowledge structure maps from enterprise ontology and to analyse Bpr steps, and it facilitated Bpr in some ways. Before explaining the proposed method some explanations about enterprise ontology is needed: One of new usages of ontology is using it for describing elements, concepts, and structures of enterprise and business. Using ontology in enterprise area, caused enterprise ontology. Enterprise ontology Is a collection of terms and definitions relevant to business enterprises. It was developed as part of the Enterprise Project a collaborative effort to provide a method and a computer toolset for enterprise modelling. A goal of the Enterprise Project is to provide a computer based toolset which will help capture aspects of a business and analyse these to identify and compare options for the meeting the business requirements. purpose is to understand the essence of the construction and operation of complex systems, more specifically, of enterprises. In this paper DEMO (Figure 1.) methodology is used for implementing enterprise ontology.

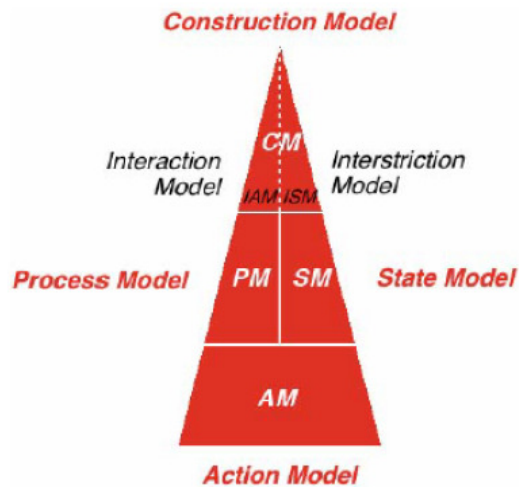

Figure 1. DEMO models

## 2.4. The Demo models

The Construction Model (CM): The CM specifies the construction of the organization, more specifically and according to [10], the CM of an organization specifies its composition, its environment, and its structure. The composition and environment of the organization are considered to be both a set of actor roles. The CM is composed by two other models, namely the interaction model (IAM), which shows the active influences between actor roles, and the interstriction model (ISM), which shows the passive influences between actor roles[12]. The IAM is composed by the Transaction Result Table (TRT) and the Actor Transaction Diagram (ATD). The TRT identifies the transactions, and the corresponding result for each transaction. The ATD represents the transactions and the participating actors, and the relations between transactions and actors. The ISM is composed by the Actor Bank Diagram (ABD) and the Bank Contents Table (BCT). The BCT identifies the production and coordination banks, and the ABD inserts these banks into the ATD, which adding extra information links results in the Organization Construction Diagram (OCD)[11].

The Process Model (PM): The PM of an organization is, according to [10], the specification of the state space and the transition space of the C-world. This means that the PM specifies the transaction pattern, for every transaction defined in the CM, and also the relations between transactions, and the participating actors. The PM is expressed by the Process Structure Diagram (PSD) and the Information Use Table (IUT). The PSD specifies the process steps for each transaction and the relations within composed transactions, and also the participating actors. The IUT identifies the object classes, the fact types and the result types, and relate them with the process steps[12].

The Action Model (AM). The AM consists of a set of action rules, which serve as guidelines for an actor. So these rules serve to define the actions the actors should take, but sometimes the actor might need to deviate from an action rule, and ultimately the responsibility, for his actions, it's his. According to[10] the AM is the basis of the other aspect models since it contains all information that is also contained in the CM, PM and SM[11].

The State Model (SM): According to [10], the SM specifies the state space : the object classes and fact types, the result types, and the ontological coexistence rules, that are contained in the AM. The SM may be viewed as the detailing of the contents of the information banks (coordination and production banks), which are part of the CM. The SM is expressed by the Object Fact Diagram (OFD) and the Object Property List (OPL). The OFD represents the object classes, identified in the PM by the IUT, and their relations. The OPL lists the object classes and their respective properties. The OFD is based on WOSL[10, 11].

The process of operation of the DEMO methodology, in terms of producing the aspect models is anticlockwise, which starts with the IAM of the CM. In order to start the IAM it is necessary a list of identified transactions and the participating actor roles, as well as the identification of the boundary of the enterprise. After the IAM, expressed by the ATD and the TRT, follows the PM, expressed by the PSD and the IUT. Next is the AM, which is expressed in a pseudo-algorithmic language, followed by the SM, expressed in the OFD and the OPL. Finally the CM is finished with the ISM, composed by the ABD and the BCT, which result in the OCD [10, 11]. Here all of the models are not needed and some of them only will be used.

Six main steps of Bpr in different approaches can be seen:

1. Preparing: this steps begins before starting doing Bpr and it's goal is to find out whether or not the organization needs Bpr, readiness of staffs and managers for Bpr, problems and challenges.
2. Analysing current system: this step shows how system really works, what are the business processes and who does which process.
3. Choosing the processes that should be reengineered: after analyzing according to Bpr budget and priorities this step shows the concentration of Bpr on processes.
4. Redesigning the processes : in this step we design the state that system should be in, in order to analysing it before implementation
5. Implementation: after approving of redesigned processes implementing them begins.
6. Evaluation and Continuous improvement: with evaluation it is possible to measure the improves caused by Bpr. And ending of Bpr is the beginning of continuous improvement for achieving Bpr goals for organization.

Of these six steps, doing the first step depends on the kind of the organization, and it is different in various places, fourth step depends on second and third steps and can't have fixed rules, there are some methods for doing it that one may choose one of them, implementation step is whether radical or first some processes will be implemented and after a while all processes would be implemented. Concentration of this paper is on second and third steps and proposed method is based on improving these steps.

## 3. Proposed Methodology

In this section the proposed method for bpr is introduced, this method consists of six steps and it's goal is to present an approach for Bpr based on enterprise ontology.

First step consist checking readiness of organization, staffs and managers for change, introducing importance of changes to staffs and managers, introduction of Bpr to them and acquiring their opinions on Bpr. To do so, in some meetings, first manager's opinions on change would be acquired, then Bpr will be discussed and introduced to staffs and mangers and at last their ideas and opinions will be gathered.

Second step, that is analysing the current system, enterprise ontology will be used for modeling and then analysing the processes, here DEMO enterprise ontology will be used. First by direct observing the performing of organization's processes, from begin to end, in one week period, and interviewing staffs and managers and customers about what they do and how they and others do their work, processes, processes results and performers of processes will be identified and listed in Transaction Result Table (TRT). Then Actor Transaction Diagram (ATD) will be made by identifying actors and their transactions. After that banks that processes and actors use it's information will be added to ATD, result is Construction model (Figure 2). Then PSD and IUT (See Figure 4) which in order are: identifying each transaction's steps and it's inner relations, and information's on these relation will be prepared result will be enterprise ontology.

Then analysing the enterprise ontology takes place in two steps, first Bpr team will do it, from process model, processes that does not work, or work poor, or can be improved will be identified and listed with suggestions for reforming them, then from construction model and IUT inner relations between processes themselves, processes with environment and staffs will be analysed, so Bpr team can have a comprehensive insight and view on organization .then gained information will be discussed with staffs and managers and meanwhile enterprise ontology will be introduced to them in order to give them a better view on processes and organization, then their opinions will be acquired.

Third step is choosing processes to reengineer, problem is that some of processes will be reengineered and after implementing them, they don't match with non- reengineered processes, and the organization fails to perform correctly[13], so the suggestion is after choosing processes to reengineer, by checking the construction and process model, processes which have the most connections with chosen ones would be also reengineered, in order to avoid inconsistency after implementation[9]. This reduces failure rate of Bpr. Remained processes will be improved in last step or if reforming them is easy in next step.

Fourth step is redesigning the chosen processes, best and most suitable ways of redesigning processes regarding resources and budget and Bpr Budget will be chosen. These ways of redesigning comes from analysing rival organization or from Bpr team suggestions or staffs and managers suggestions and nowadays in most cases consists of using IT for doing the processes, but it should not be just replacing old processes with IT, rather while using IT, processes must be changed and improved. In this step the chosen processes must be modeled and simulated in order to analyse and correct them before trying to implement them, and reduce the costs of implementation. It is strongly suggested and emphasized that in this step before redesigning, future changes in market and organization must be considered and best ways for reengineering must be used in order to avoid reengineering in upcoming years.

International Journal of Software Engineering & Applications (IJSEA), Vol.6, No.1, January 2015

Fifth step is implementing redesigned processes, new equipment will be installed, new systems will begin to work, new processes will be implemented, and organization will start working with them, meanwhile staffs will be educated for new works. Obtaining results will takes time, so result will show themselves not very fast.

Sixth step is evaluating the results, based reduction on times and costs of performing the processes, while evaluating and after that processes will be revised and improved continuously. Figure 3 shows the whole process .

### 3.1 Case Study

For implementing the proposed methodology a distributing segments of car's body company in city of Kerman of Iran is used, this company buys segments of car's body from factories and distribute them in city Kerman among car's accessory shops. Company consists of four departments: selling and marketing, accounting, management and storing. Purpose is doing bpr on its processes. Because it will take over a year for preparing results of implementing Bpr on the company and lack of time and money for implementation, the implementation and continuous improvement steps will be done later by the company itself and in here we compare and analyse by computer simulations.

1) At first and in first step of bpr, some presentation and introduction were done in a meeting with managers and staffs, and they declared their opinions and their readiness for starting bpr.
2) second step that is analysing the current situation of the system was done by directly observing processes of company in a specific time period and interviewing with staffs and managers and finally modelling the enterprise ontology in order to gain a better understanding of the company and its inner and outer relations. Stages of processes was in this form: after verbal marketing customers were identified and aware of company's business, next customers contact with company and selling department and order the available items, then the accounting department checks the customer's history for determining way of paying (check/cash) and make the invoice, then customer pays the invoice and informs the accounting, the paid invoice will be sent to store department to prepare it and delivering it to the customer, to order for store, seller checks the stocks for unavailable items, then he orders for store after approval of the manager and accounting department, then factory sends the invoice that will be paid after approval of manager and accounting department, then factory delivers the order to the store, in store after matching the order with invoice, cargo will be unloaded and stored. In this step many of big and small problems were identified that some the most important of them are: wrong organizational chart, unprepared and poor educated staffs, contradiction between staffs abilities and their job, dependency of whole company on manager, undefined inner rules and work hours, limited sources for delivering, uncategorized customers, lack of control and analyse on the market, late update on stocks and returned items.
3) after analysing the company and the enterprise ontology, two main process, selling and storing were choose to reengineer, and after analysing the inner relations of these two processes, the accounting department were also selected for redesign. Also a "market controlling and auditing" department were added to the company for continuous analysing of the market and customers
4) Three chosen processes were reengineered and suggestions were made for other processes. For selling, an internet website were proposed that customers can visit it and order what they need, then online orders will be sent to selling and accounting departments, selling department check the stocks in store and accounting will check the customer's history in database and determines way of paying for customer, and finally make and send him the invoice and customer will receive the order after paying the invoice. This way of selling will reduce amounts of time wasted while phone selling and reduce unnecessary communications between selling and accounting department.

For storing department using RFID and two readers suggested to facilitate listing the ingoing stocks to store and checking available stock for selling and also prevent outgoing of stock without permission and a lot of time spending on listing the stock will be saved. also a separate delivery and distributing department with vehicle and driver will be added to storing department to improve delivery system.

in accounting department am accounting software will replace the current system to improve and facilitate managing input and output invoices, gathering all customers data, and controlling all financial transactions. Market controlling department will be added to the company in order to finding better and cheaper opportunities like distributers and importers, controlling changes in market, informing changes in demands and brands, identifying progressing and falling customers, suggesting policy for behaving and trading with each customer, categorizing customers for better serving and increasing customer satisfaction. After reengineering, other processes will also face suggestions for their problems, like: changing organizational chart, reducing manager power and responsibilities in order to give it to the manager of each department for not relying on only one person, categorizing region of customers for better serving, educating storing workers and increasing storing quality of storing stock for reducing damages of them and returning of them

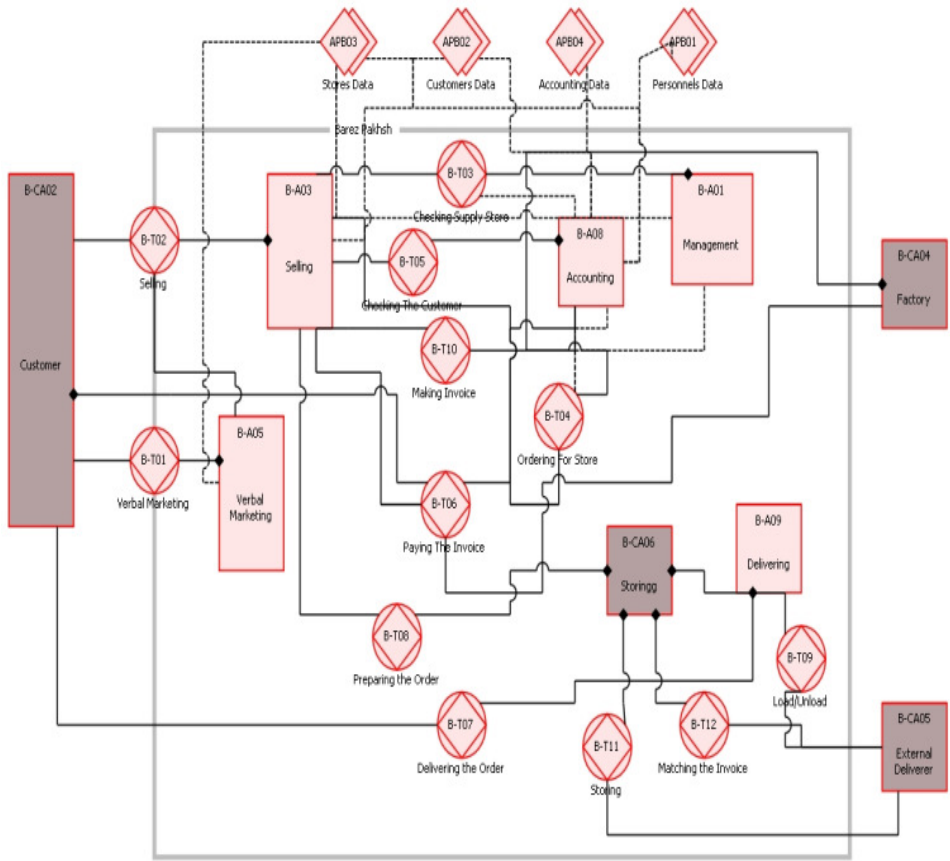

Figure 2. Construction Model of company

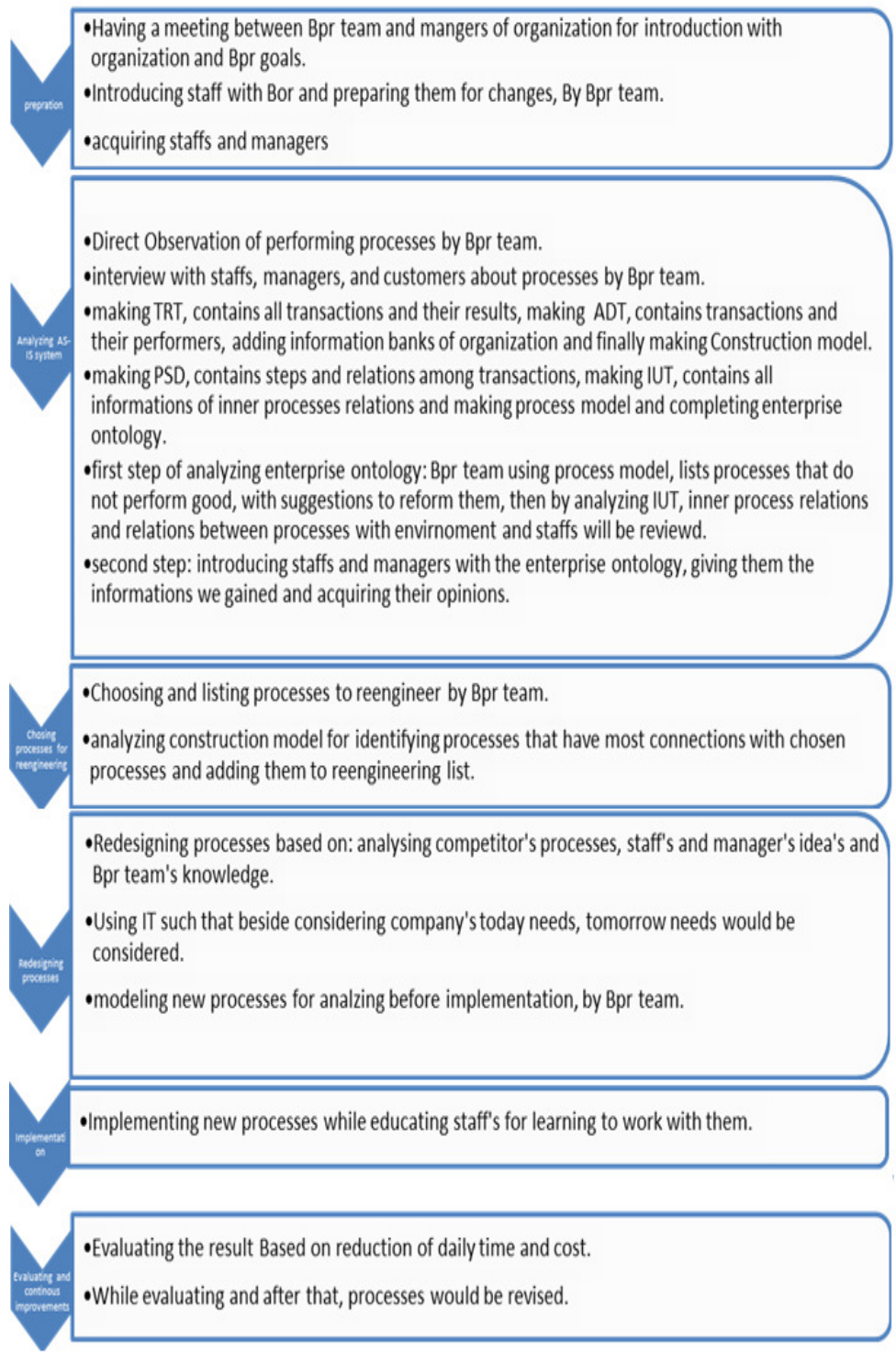

Figure 3. Proposed methodology

International Journal of Software Engineering & Applications (IJSEA), Vol.6, No.1, January 2015

## 3.2 Evaluation

For Evaluating, needed time and cost for performing company's processes before and after reengineering were measured, and compared. With tools like "Simulation Experiment" and "Report Analyse Time/Cost" in ARIS, AS-IS (Figure 5) and TO-BE (Figure 6) models of company were simulated and compared.

Figure 4. IUT

Time and cost of performing processes before reengineering were collected from company data's and analysing the current system by Bpr tram. Time unit is minute and cost unit is Euro. For calculating time and cost of processes after reengineering, for unchanged processes previous data's were used and for reengineered processes, experimental performing of those processes were used. Simulation in software for the two models were done in 6 months, with 6 workdays and 8 hours a day with 3 replications that based on number of performing each process in each model while simulating, the average result for each workday are presented in figure 8. In figure 5,6 parts of simulating result of each model is presented. Reduction in daily cost were 42% and in daily

33

time were 41%, because measuring the sale rate in TO-BE model depends on market situation, Store capacity, and order rate, without fully implementing and analysing the system in a one year period is not possible, therefore only simulating the cost and time in models were performed. For comparing this method with another method, in another branch of this company that has same processes, current situation of system has been modeled and reengineered with another bpr method [7]. After simulating, result was reduction in time and cost in sequence 25 and 23%. Figure 8 shows some of it steps. At last a comparison between proposed methodology and other methodologies based on matrix method and with factors from [14] and information in[15] presents.

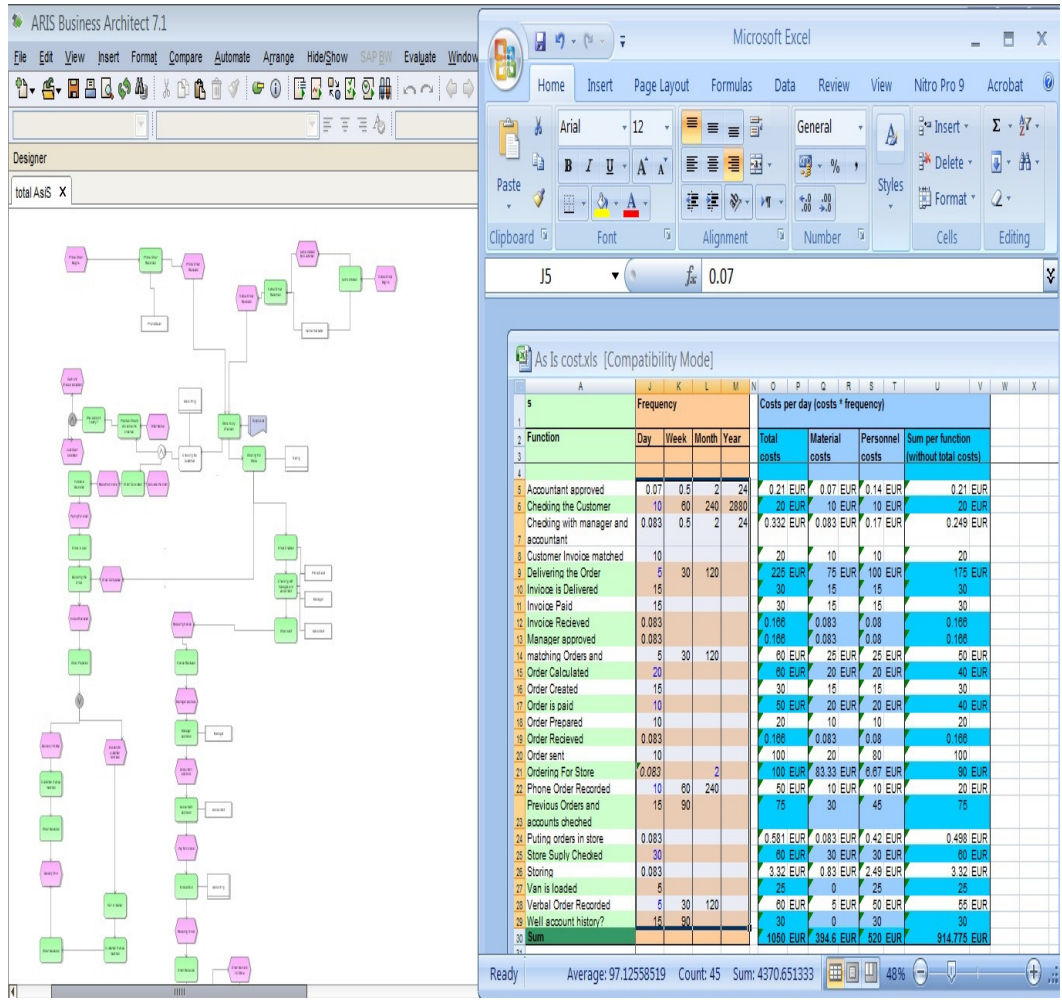

Figure 5. Analyzing AS-IS system

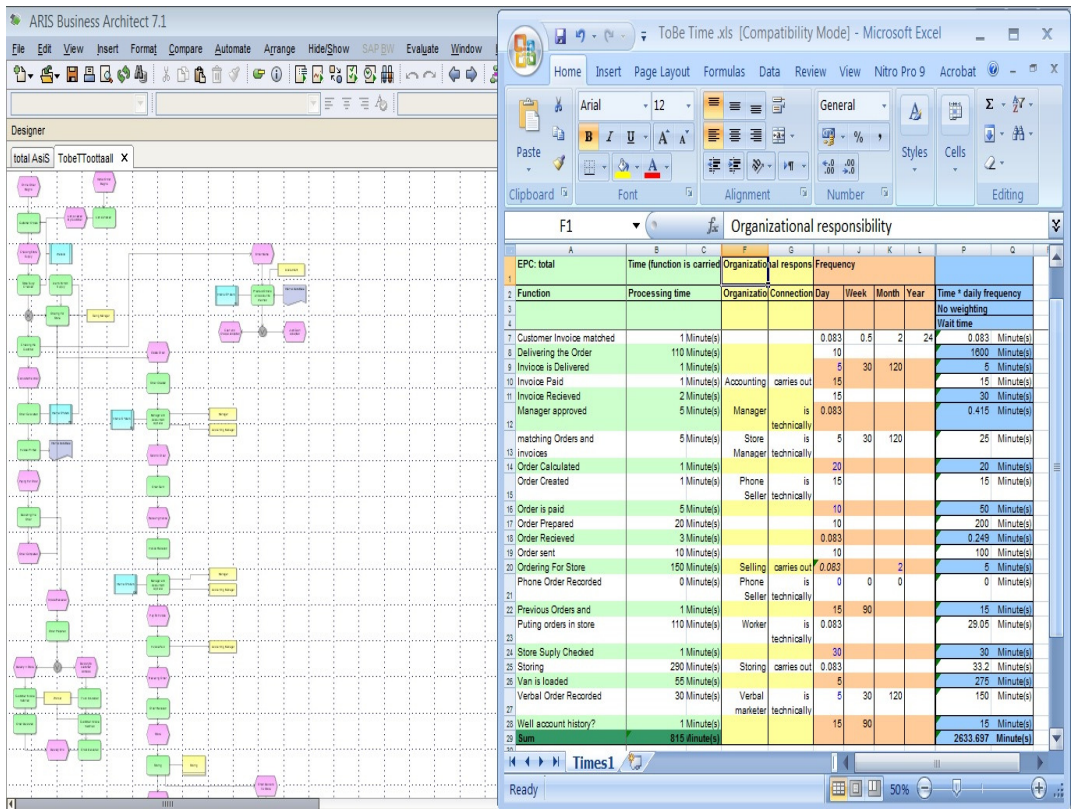

Figure 6. Analyzing TO-BE system

International Journal of Software Engineering & Applications (IJSEA), Vol.6, No.1, January 2015

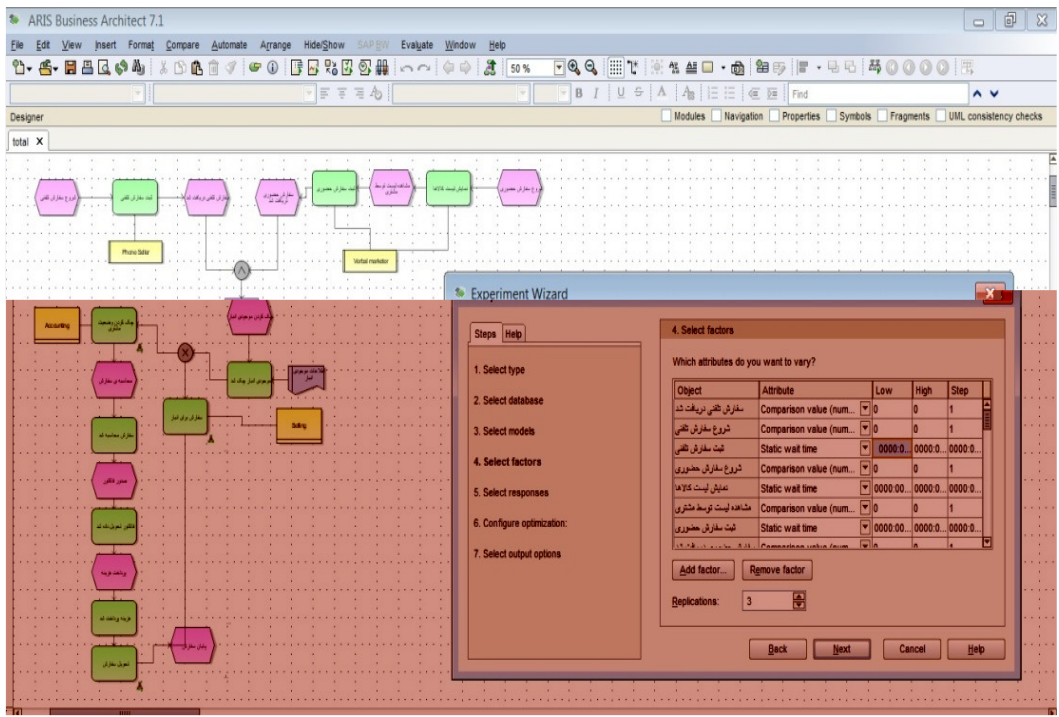

Figure 7. Analyzing compared company

International Journal of Software Engineering & Applications (IJSEA), Vol.6, No.1, January 2015

| Function | Total AS-IS costs | Time * daily frequency AS IS time | | Total ToBe costs | Time * daily frequency To Be time | |
|---|---|---|---|---|---|---|
| Accountant approved | 0.21 EUR | 4.98 | Minute(s) | 0.07 EUR | 0.7 | Minute(s) |
| Checking the Customer | 20 EUR | 200 | Minute(s) | 10 EUR | 20 | Minute(s) |
| Checking with manager and accountant | 0.332 EUR | 60 | Minute(s) | 0.083 EUR | 0.083 | Minute(s) |
| Customer Invoice matched | 20 | 50 | Minute(s) | 10 | 1 | Minute(s) |
| Delivering the Order | 225 EUR | 900 | Minute(s) | 200 EUR | 1600 | Minute(s) |
| Invioce is Delivered | 30 | 75 | Minute(s) | 15 | 5 | Minute(s) |
| Invoice Paid | 30 | 150 | Minute(s) | 15 | 15 | Minute(s) |
| Invoice Recieved | 0.166 | 0.415 | Minute(s) | 0.083 | 30 | Minute(s) |
| Manager approved | 0.186 | 0.83 | Minute(s) | 0.083 | 0.415 | Minute(s) |
| matching Orders and invoices | 60 EUR | 900 | Minute(s) | 10 EUR | 25 | Minute(s) |
| Order Calculated | 60 EUR | 100 | Minute(s) | 20 EUR | 25 | Minute(s) |
| Order Created | 30 | 75 | Minute(s) | 0 | 15 | Minute(s) |
| Order is paid | 50 EUR | 600 | Minute(s) | 50 EUR | 50 | Minute(s) |
| Order Prepared | 20 | 300 | Minute(s) | 10 | 200 | Minute(s) |
| Order Recieved | 0.166 | 0.83 | Minute(s) | 0.083 | 0.249 | Minute(s) |
| Order sent | 100 | 300 | Minute(s) | 90 | 100 | Minute(s) |
| Ordering For Store | 100 EUR | 10 | Minute(s) | 100 EUR | 5 | Minute(s) |
| Phone Order Recorded | 50 EUR | 100 | Minute(s) | 0 EUR | 15 | Minute(s) |
| Previous Orders and accounts cheched | 75 | 150 | Minute(s) | 0 | 15 | Minute(s) |
| Puting orders in store | 0.581 EUR | 29.88 | Minute(s) | 0.581 EUR | 29.5 | Minute(s) |
| Store Suply Checked | 60 EUR | 60 | Minute(s) | 0 EUR | 30 | Minute(s) |
| Storing | 3.32 EUR | 34.86 | Minute(s) | 1.245 EUR | 33.2 | Minute(s) |
| Van is loaded | 25 | 300 | Minute(s) | 25 | 275 | Minute(s) |
| Verbal Order Recorded | 60 EUR | 150 | Minute(s) | 60 EUR | 150 | Minute(s) |
| Well account history? | 30 | 75 | Minute(s) | 0 | 15 | Minute(s) |
| **Sum** | **1049.941 EUR** | **4566.8** | **Minute(s)** | **617.23 EUR** | **2654.4** | **Minute(s)** |

Figure 8. Comparing AS-IS and TO-BE

### 3.3 Results and Conclusions

Using enterprise ontology for modeling the system, causes better modeling the essence of the enterprise. Internal and external relations are fully determined, better analysis will be gained and causes improving the understanding the current system and therefore better finding of problems in system. Redesigning the processes with most connections with the chosen processes to reengineer, causes improving the performance of system and Bpr. Also using simulation in redesigning step and before implementation would cause reduction in implementation's costs and finally all of these cause reduction in failure rate of Bpr , Cost and time of performing the processes. Table 1. compares proposed methodology with other methodologies.

International Journal of Software Engineering & Applications (IJSEA), Vol.6, No.1, January 2015

Table 1. Comparing with other methods

| | Hammer and champy | Davenport | Kodak | Manganelli and Klein | Sanjay | Eftekhari and Akhavan | Proposed Method |
|---|---|---|---|---|---|---|---|
| Emphasize on analyzing current system | Low | High | Average | Average | High | Average | High |
| IT Role | Average | High | High | High | High | High | High |
| Adoptability with different organization | Average | Average | High | High | Average | Average | Average |
| Adoption with changes | Average | Average | High | Average | Average | High | Average |
| Emphasize on consistency | Low | Low | Average | Low | Average | Average | High |
| Using staff's knowledge | Average | Average | Low | High | High | Average | High |
| Emphasize on organization's structure | Low | Average | Average | Low | Low | Average | High |
| Limits | Loss of analyzing current system- Having no attention to consistency | Long performing time | Low attention to analyzing current system | Having no attention to consistency | Having low attention to consistency | Having Low attention to consistency | Complexity of analyzing current system |